\documentclass[nocompress]{spie}  

\usepackage{amsmath,amsfonts,amssymb}
\usepackage{graphicx}
\usepackage[colorlinks=true, allcolors=blue]{hyperref}
\usepackage{epstopdf}
\usepackage{wrapfig}
\usepackage[ruled,linesnumbered]{algorithm2e}

\begin{document}
\title{Tapering Analysis of Airways with Bronchiectasis}

\author[a]{Kin Quan}
\author[b]{Rebecca J. Shipley}
\author[a]{Ryutaro Tanno}
\author[c]{Graeme McPhillips}
\author[a,f]{Vasileios Vavourakis}
\author[e]{David Edwards}
\author[a]{Joseph Jacob}
\author[d]{John R. Hurst}
\author[a]{David J. Hawkes}
\affil[a]{Centre for Medical Image Computing, University College London, Gower Street, London, UK}
\affil[b]{Dept of Mechanical Engineering, University College London, Gower Street, London, UK}
\affil[c]{Dept of Computer Science, University College London, Gower Street, London, UK}
\affil[d]{UCL Respiratory, University College London, Gower Street, London, UK}
\affil[e]{Radiology, The Royal Free Hospital, Pond Street, London, UK}
\affil[f]{Dept of Mechanical and Manufacturing Engineering, University of Cyprus, University Avenue, Nicosia, Cyprus}

\authorinfo{Send correspondence to Kin Quan \\Kin Quan: E-mail: kin.quan.10@ucl.ac.uk}

\pagestyle{empty} 
\setcounter{page}{1} 

\maketitle

\begin{abstract}
Bronchiectasis is the permanent dilation of airways. Patients with the disease can suffer recurrent exacerbations, reducing their quality of life. The gold standard to diagnose and monitor bronchiectasis is accomplished by inspection of chest computed tomography (CT) scans. A clinician examines the broncho-arterial ratio to determine if an airway is brochiectatic. The visual analysis assumes the blood vessel diameter remains constant, although this assumption is disputed in the literature. We propose a simple measurement of tapering along the airways to diagnose and monitor bronchiectasis. To this end, we constructed a pipeline to measure the cross-sectional area along the airways at contiguous intervals, starting from the carina to the most distal point observable. Using a phantom with calibrated 3D printed structures, the precision and accuracy of our algorithm extends to the sub voxel level. The tapering measurement is robust to bifurcations along the airway and was applied to chest CT images acquired in clinical practice. The result is a statistical difference in tapering rate between airways with bronchiectasis and controls.\\

Our code is available at \url{https://github.com/quan14/AirwayTaperingInCT}\\

The following manuscript was previously submitted for SPIE Medical Imaging, 2018, Houston, Texas, United States. The citation is as follows: K. Quan et al., "Tapering analysis of airways with bronchiectasis," Proc. SPIE 10574, Medical Imaging 2018: Image Processing, 105742G (2 March 2018); \url{https://doi.org/10.1117/12.2292306}.

\end{abstract}


\section{INTRODUCTION}
\label{sec:intro}  

The British Thorax Society defines bronchiectasis as permanent dilatation and damage of the airways\cite{Pasteur2010}. In clinical practice, visual inspection is used to diagnose and assess the severity of bronchiectasis. A radiologist will diagnose an airway as bronchiectatic if its diameter is bigger than its accompanying pulmonary artery\cite{Pasteur2010}. These observations are placed in radiological scoring systems for bronchiectasis like the Bhalla score\cite{Bhalla1991}. Recent works showed these radiological scoring systems combined with other clinical data are effective at predicting hospitalisation and mortality\cite{Chalmers2014}.

However, the process of diagnosing bronchiectasis suffers from two disadvantages. First of all, established radiological scoring systems are complicated, requires various inputs and are based on lobes; not individual airways\cite{DeJong2007}. Second, the entire process assumes the diameter of blood vessel remains constant whereas the literature has suggested that the size of vessel can change depending on age\cite{Matsuoka2003}, altitude\cite{Kim1997}, and smoker status\cite{Diaz2017}. 

We believe acquiring detailed geometric information on the airways may enable quantification of bronchiectasis. To this end, we measured the tapering of each airway i.e. the rate of change of cross sectional area along its centreline. The proposed measure is independent of the accompanying blood vessel and lobe. We hypothesise that the taper of bronchiectatic airways are lower in magnitude compared to radiological healthy airways. The rationale is that bronchiectatic airways will retain a larger cross sectional area along the centreline.

It has been observed that tapering are different in airways affected by bronchiectasis compared to healthy patients\cite{Odry2008b}. In a study by Weinheimer et al.\cite{Weinheimer2017} they proposed a graphical based tapering measurement to monitor children with bronchiectasis. However, there has been little work to quantify the differences in tapering between diseased airways and airways that are radiologically normal.

This study presents a methodology used to measure tapering of airways on clinically acquired chest CT images. The algorithm measures the arc length and cross sectional area at contiguous intervals along the airway. Its precision and accuracy were validated using a physical phantom containing calibrated 3D printed structures. Finally, comparisons were made to bronchiectatic airways defined by a radiologist (JJ).

\section{METHODS AND MATERIALS}

\subsection{Image Pre-processing}
CT images were acquired from 9 patients after obtaining patient consent at the Royal Free Hospital, London. The voxel size varies from 0.63-0.80mm in plane and 0.80-1.5mm slice thickness. In each CT dataset, a radiologist (JJ) identified a group of bronchiectatic airways and a group of healthy airways. The proposed pipeline requires two inputs: the distal point of the airway and a complete segmentation of the airway of interest.

\subsubsection{Distal Point}
For each airway, the radiologist manually identified the most distal point of the airway that was visible on the CT image. The protocol is as follows; when a suitable airway is chosen, the radiologist manually tracks the airway towards the most distal point that is visible in the CT image. The tracking was performed entirely on the axial slice. Once the distal point has been reached, the radiologist marked the point with a single voxel via a mouse cursor. In the case where the airways were dilated at the distal point, the radiologist marked the point where it is judged to be the end of the centreline.

\subsubsection{Airway Segmentation}
We performed the binary airway segmentation using a third-party software developed by Rikxoort et al\cite{Rikxoort2009}. The airway segmentation is based on an adaptive region growing algorithm.

The output is a binary segmentation of the airway tree. Every binary image was visually validated by the author. Any leaks in the segmented image were removed. Furthermore, in the case where the airway segmentation had not reached the distal point, the author (KQ) interactivity extended the segmentation to the distal point.

\subsection{Pipeline}
\begin{figure}
	\centering
		\includegraphics[trim={0 0 0 0},clip,width=1.0\textwidth]{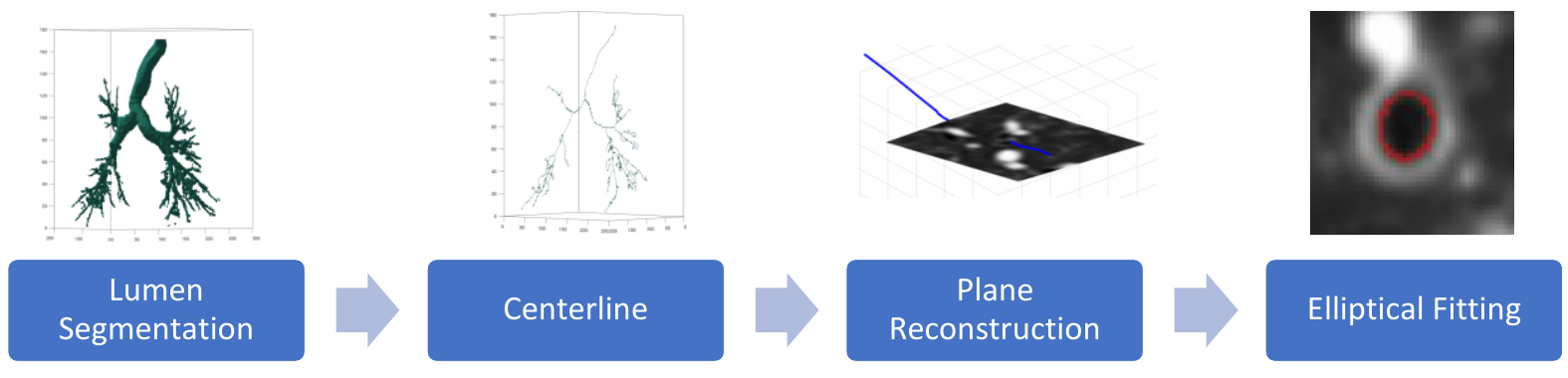}
	\caption{A summary and visualisation of the processes of the pipeline.}
	\label{Tappering_pipline}
\end{figure}

The pipeline consists of the following steps; generating the centreline, construction of an image plane, and elliptical fitting. The steps is summarised in Figure \ref{Tappering_pipline}. For each airway, the aim was to generate two measurements: a series of cross-sectional areas from the carina to the most distal point visible on the image together with the arc length from the carina to the respective points.

\subsubsection{Centreline}
We implemented a curve thinning method developed by Palagyi et al\cite{Palagyi2006}. To summarise, the algorithm iteratively removes voxels on the boundary of the binary image. If the removed voxel causes a change in the topology of the image then it is placed back into the image. The method requires the endpoints of the centreline to be demarcated. These points correspond to the start of the trachea and the distal points. The start of the trachea was computed by taking the distance transform of the airway segmentation, $D$ and applying Algorithm \ref{Start_finder}. The algorithm assumes the local maximum point of the distance transform corresponds to a centreline point in a tubular object \cite{Siddiqi2008,Aylward2002}. From the generated centreline, we removed the centreline points corresponding to the trachea.

\newcommand{\argmax}{\operatorname{Argmax}}
\begin{algorithm}[ht]
	\SetAlgoLined
	\KwData{$D(i)$ Distance image on the $i$th axial slice}
	\KwResult{$x_{s}$ Start point of trachea}
	$i\leftarrow $ First slice at the top of the trachea.\\
	\While{$\max{D(i)} < \max{D(i+1)}$}{
		$i = i + 1$
	}
	$x_{s} = \argmax{D(i)}$
	\caption{Locating the start of centreline on the trachea}\label{Start_finder}
\end{algorithm}

The next step is to separate the centreline tree into individual airway centrelines and remove any false branches. To this end, a breadth first search algorithm\cite{Cormen2009}  was performed to find the unique paths of the centreline for each airway. Starting from the carina, we iteratively find the next set of daughter branches. When a distal point was found, then the path leading to the distal point is saved. The algorithm continues until all distal points are identified. The final output is a set of 26 connected voxel paths to each corresponding distal point. By performing a breadth first search algorithm, the false branches are automatically removed.

The centreline was corrected for any discretization errors, a process known as re-centring\cite{Grelard2017}. In this study, a five-point smoothing approach was adopted on each point on the centreline. Next, we require a continuous model to compute the tangent along the airway. To this end, we fitted a cubic spline to every point on the centreline. This method shares similarity to the methodology proposed by Irving et al\cite{Irving2014}.

\subsubsection{Plane Reconstruction}

The aim of the plane reconstruction was to generate an image plane that is perpendicular to the tangential direction of the vessel. The image generated was used to compute the cross sectional area of the lumen. For our pipeline, two image planes are computed. One plane from the CT image and another from the airway segmentation image. Both planes are located at the same physical point.

At each 0.25 parametric interval on the spline, we compute the tangent vector. Next, we find a set of basis for the orthogonal plane by Algorithm \ref{Normal_basis}. Furthermore, at each point of the parametric interval, we also compute the arc length from the start of the carina to the sampled location.

\begin{algorithm}[ht]
\SetAlgoLined
\KwData{$t$ Unit tangent vector of the spline}
\KwResult{$v_{1}$,$v_{2}$ Basis of the orthogonal plane}
$a \leftarrow $ Arbitrary vector such that $a$ and $t$ are not collinear \\
$v_{1} = \frac{a \times t}{|a \times t|}$ \\
$v_{2} = v_{1} \times t$
\caption{Constructing the basis for the plane reconstruction, adapted from Shirley and Marschner\cite{Shirley2009}.}\label{Normal_basis}
\end{algorithm}

The computed basis vectors are used for generating the points on both planes. The points generated have a pixel size of 0.3mm and dimensions 40mm by 40mm. The parameters were chosen to ensure the entire lumen was in the field of view. The intensities of the two planes were computed by cubic interpolation.

\subsubsection{Elliptical Fitting}


To compute the cross-sectional area of the airway, an ellipse was fitted to the boundary of the lumen. To this end, the lumen boundary was found through a ray casting method using the two plane images computed from the previous section. The method is known as the Edge-Cued Segmentation-Limited Forward Half Width Maximum (FWHM\textsubscript{ESL}), developed by Kiraly et al\cite{Kiraly2005}. In belief, 50 rays are cast out in a radial direction, from the centre of the plane. Each ray samples the intensity of the two planes at a fifth of a pixel via linear interpolation. Thus each ray produces two 1D images, first from the binary plane $r_{b}$, and second from the CT plane $r_{c}$. We then apply Algorithm \ref{ray_pt} to find boundary point $l$.

\begin{algorithm}[ht]
\SetAlgoLined
\KwData{The rays: $r_{b}:[0,p] \rightarrow \mathbb{R}_{[0,1]}$, $r_{c}:[0,p] \rightarrow \mathbb{R}$ where $p$ is the length from the centre to the border of the plane.}
\KwResult{The position of the lumen edge, $l$.}
$s \leftarrow $ The first index of the ray such that $r_{b}(s)<0.5$ \\
$I_{max} \leftarrow $ Local maximum intensity in $r_{c}$ nearest to s\\
$x_{max} \leftarrow $ The index such that $r_{c}(x_{max}) = I_{max}$\\
$I_{min} \leftarrow $ Minimum intensity in $r_{c}$ from $0$ to $x_{max}$ \\
$x_{min} \leftarrow $ The index such that $r_{c}(x_{min}) = I_{min}$ \\
$l \leftarrow $ The index such that $r_{c}(l) = (I_{max} + I_{min}) \times 0.5$ and $l \in[x_{min} , x_{max}]$
\caption{Summary of the FWHM\textsubscript{ESL}, adapted from Kiraly et al\cite{Kiraly2005}. The purpose of the algorithm is to find the point of the ray which crosses the lumen.}\label{ray_pt}
\end{algorithm}

The output of the ray casting is a set of 2D points $(x,y)$, corresponding to the edge of the lumen. From these points, an ellipse is fitted based on the least square principle, developed and implemented in Matlab by Fitzgibbon et al\cite{Fitzgibbon1996}. The cross-sectional area was computed from the fitted ellipse. 

\subsubsection{Tapering Measurement} \label{T_Measurement}
\begin{figure}
	\centering
	\includegraphics[trim={0 0 0 0},clip,width=0.9\textwidth]{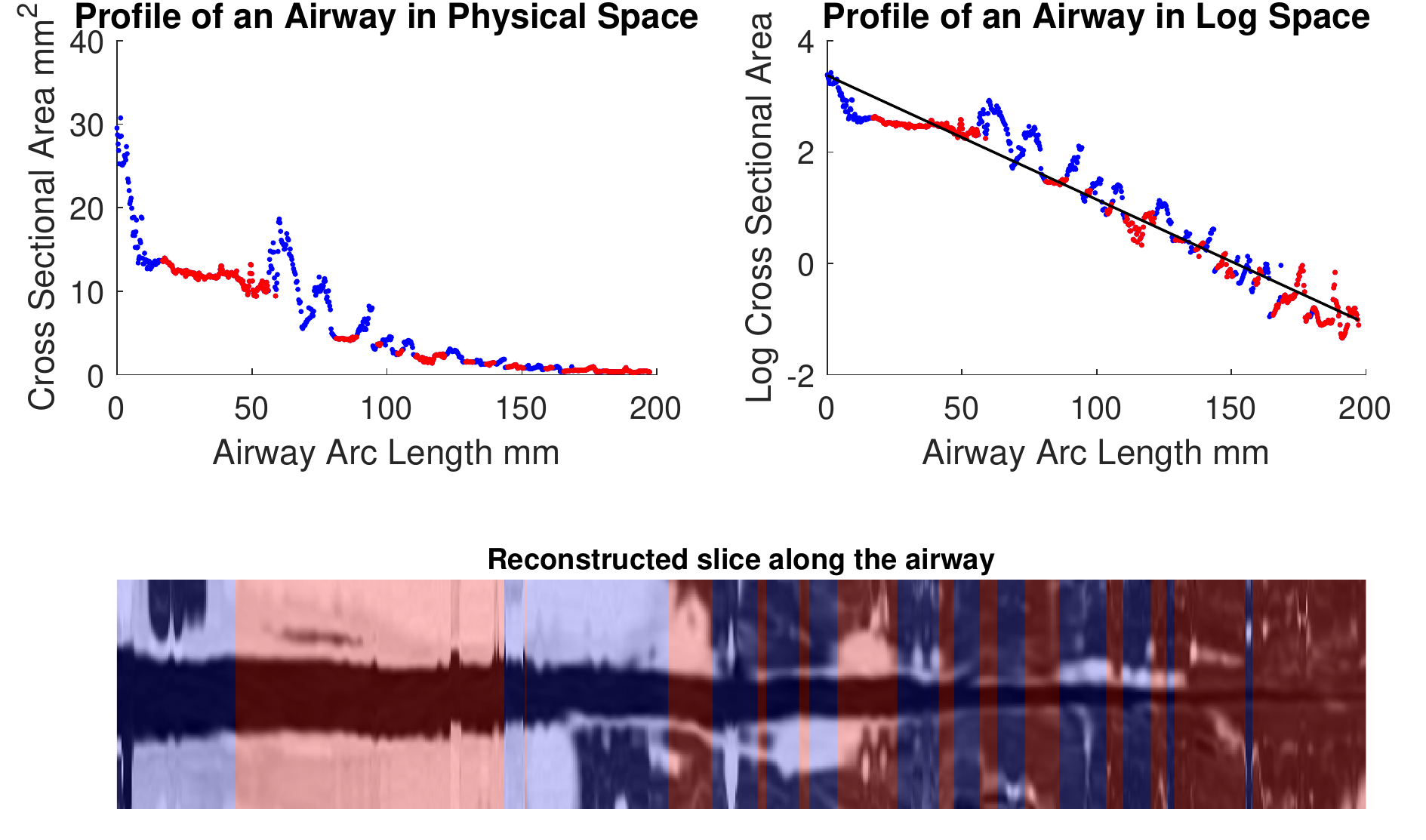}
	\caption{TOP LEFT:  A profile of cross sectional area along an airway. The blue points are regions of bifurcations and red points are where the airways are tubular. TOP RIGHT: The same profile in log space with a line of best fit. The gradient of the line is our tapering measurement. BOTTOM: The reconstructed image of the profile, the blue-shaded and red-shaded regions corresponds to bifurcating and tubular airways respectively.}
	\label{Method_ind_airway}
\end{figure}

We model the airway tapering as an exponential function between the cross-sectional area and centreline. Thus, a logarithmic transform, $\log(a)$, was applied on all cross-sectional area measurements, $a$. We defined the tapering measure as the gradient of the linear fit between the log cross-sectional area and the arc length. An example of an airway profile is displayed in Figure \ref{Method_ind_airway}.

\subsection{Bifurcation Analysis}
\begin{figure}
	\centering
		\includegraphics[trim={0 0 0 0},clip,width=0.9\textwidth]{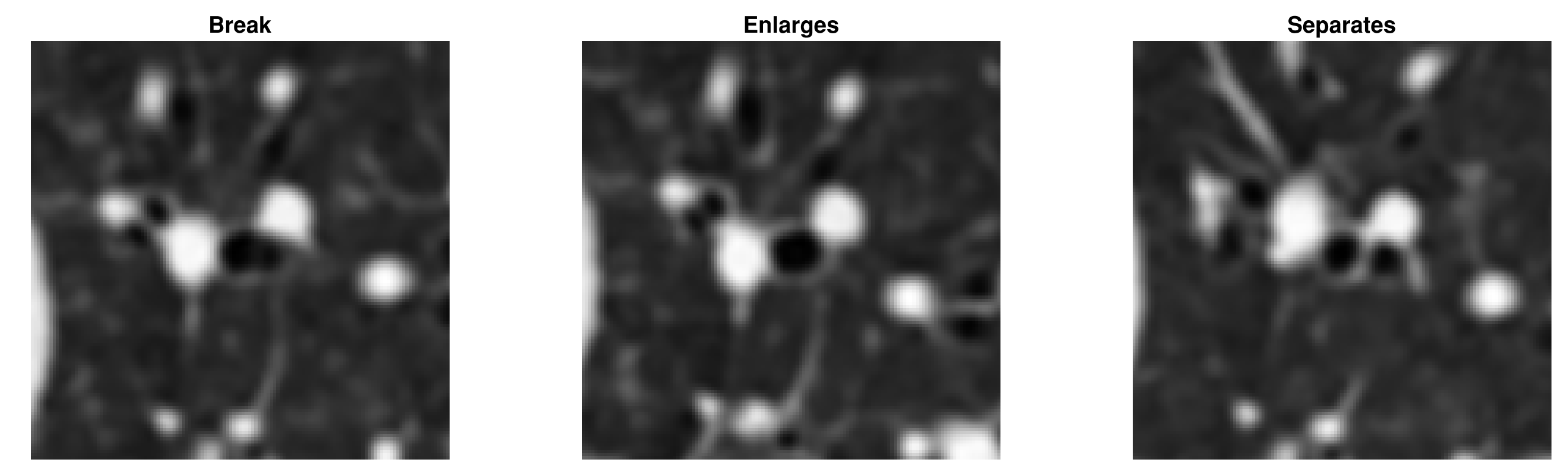}
	\caption{Examples of the stages in identifying a bifurcation region. We scroll along the airway cross-sections until a break is detected. We scroll back until the enlargement caused by the bifurcation is found. We scroll forward, marking every slice as a bifurcating region until the airway separates.}
	\label{Bi_steps}
\end{figure}

We consider how bifurcations influence the taper measurements. On the same dataset, 19 airways were selected (11 controls, 8 diseased) from 9 patients. To identify the bifurcation points, we used the output of the reconstructed plane image. The image consisted of a series of lumen cross sections correspondingly from the start of the carina to the distal point. Examples are displayed in Figure \ref{Bi_steps}. The dimension of each plane was 133 by 133 pixels.

A visual protocol was developed for identifying the bifurcation. The protocol is as follows; for each reconstructed image, we start on the plane corresponding to the start of the carina. Using the xy-plane, we scroll down and mark according to the following appearance:
\begin{enumerate}
	\item We scroll until a ``break'' is detected in the airway caused by the bifurcation.
	\item From the break point we scroll back to the point before the airway ``enlarges''.
	\item We then mark every slice as bifurcating until the airway fully ``separates''.
\end{enumerate}

\subsection{Implementation}
The majority of the pipeline was implemented using Matlab. The distance transform was performed by our in house software using the ITK library\footnote{\url{http://itk.org}, last accessed on \today}. Code from the Pulmonary Toolkit\footnote{\url{http://github.com/tomdoel/pulmonarytoolkit}, last accessed on \today} was used to generate the centreline of the airways. All manual inspection and delineation of images were performed using ITK-snap\footnote{\url{http://www.itksnap.org/}, last accessed on \today}.

\subsection{Phantom Experiment}
The aim of the phantom was to have a ground truth to assess the accuracy and precision of the pipeline. The design of the phantom was to encapsulate various morphologies of the airway lumen. To this end, the airway lumen structures were built using 3D printing.

\subsubsection{Phantom Design}
The body of the phantom was a cylindrical Perspex case, 240mm in diameter. A set of 3D printed structures was attached on the flat side of the cylinder. The remaining space was filled with rice to approximately mimic the attenuation properties of lung parenchyma. A similar material was used by Robinson et al.\cite{Robinson2009} for the same effect.

\begin{figure}[ht]
	\centering
		\includegraphics[width=1.00\textwidth]{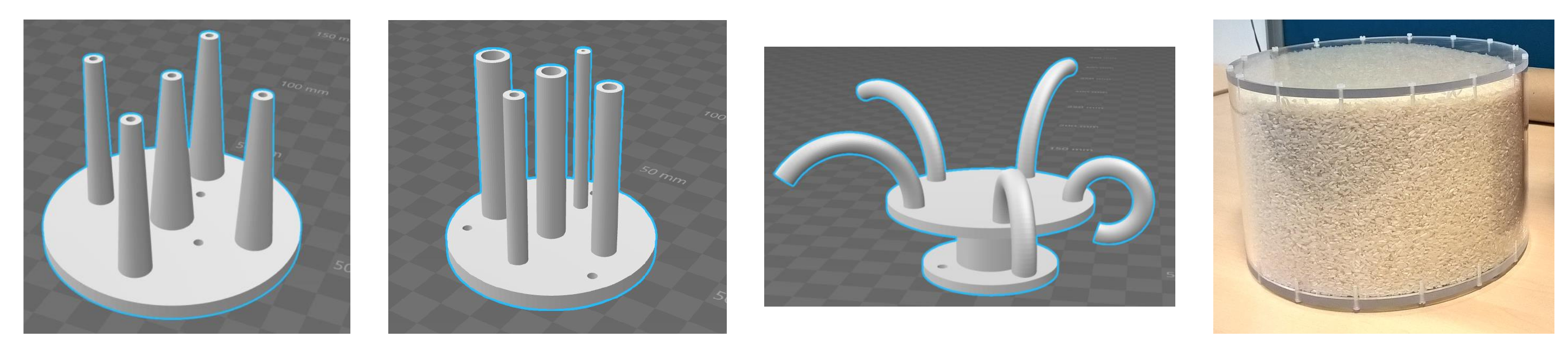}
		\caption{FAR LEFT: Different tapering lumen. CENTRE LEFT: Different lumen diameter. CENTRE RIGHT: Different lumen centreline curvatures. FAR RIGHT: Physical phantom.}
		\label{Printed_sturcts}
\end{figure}

We designed the 3D printed structures to simulate various morphologies based on the appearance of airways in CT. Three parameters were used; diameter, radius of curvature and tapering. Each structure consists of 5 tubes attached to a circular base, with wall thickness 1.7mm and length 50mm. The designs are displayed in Figure \ref{Printed_sturcts}. All structures were made using an EnvisionTEC ULTRA 3SP printer with the ABS 3SP Tough resin. The voxel resolution ranges from 0.05mm to 0.1mm. The resolution was set to the lowest possible setting. The supporting structure was set at the base of the tubes.

The diameter lumen consist of tubes of diameters: 1.1mm, 2.5mm, 3.9mm, 5.3mm, and 6.7mm. The curvature lumen consists of 2.5mm diameter tubes with its centreline radius of curvature of 30mm, 25mm, 20mm, 15mm, and 10mm. For the tapering structures, we used the linear change in diameter along the tube. All tapering tubes started at 2.5mm at the tip and diameter $d$ along the tube was calculated as:
\begin{equation}
d = 2.5 + zt,
\end{equation}
with centreline arc length $z$ and diameter gradient $t$. The diameter gradient varies with each tube, as shown on Table \ref{tapering_table}. The parameters were chosen from visual inspection of the CT image by the author (KQ) and cadaver experiments from the literature.\cite{Nikiforov1985}

\begin{table}
\centering
	\begin{tabular}{ c | c | c | c }
		Tube Number & Diameter Gradient $t$ & Start Diameter/mm & End Diameter/mm \\ \hline
		1 & 0.051 & 2.5 & 5.1 \\
		2 & 0.083 & 2.5 & 6.7 \\
		3 & 0.109 & 2.5 & 8.0 \\
		4 & 0.132 & 2.5 & 9.1 \\
		5 & 0.168 & 2.5 & 10.4 \\
	\end{tabular}
				\caption{The parameters of each tube in the tapering phantom.}
				\label{tapering_table}
\end{table}

\subsubsection{Image Acquisition and Post Processing}
The phantom was imaged in a Toshiba Aqulilion ONE CT scanner at the Royal Free Hospital. The image voxel size were 0.625mm by 0.625mm by 1mm and reconstructed using the Lung kernel. To test the pipeline, every tube was segmented using SegEM\footnote{\url{http://cmictig.cs.ucl.ac.uk/wiki/index.php/Seg_EM}, last accessed on \today.} with manual corrections. The endpoints were semi-manually demarcated using the centre of mass\cite{Jain1989} of each segmented tube.

\subsubsection{3D Printer Analysis}
The scale of the CT image is sub-millimetre, thus, we verified the precision and accuracy of the 3D printer. To this end, two 2.5mm diameter lumen was manufactured - one 3D printed and another from a lathe. The milled lumen had a tolerance of 0.05mm.

Both lumens were micro CT scanned consecutively. The scanner was a Skyscan 1172 with a voxel size of 11.0$\mu$m isotropically. The large image size made it difficult to perform any post-processing. Thus we downsampled the images to 22$\mu$m isotropically with Sinc interpolation.

To avoid computing the centreline, the lumen needed to be perpendicular to the in-plane slice. To this end, the lumen was initially semi-manually segmented with SegEM. The misaligned angles was computed through the centre of mass of the segmented lumen. The image was then rotated with the misaligned angles with Sinc interpolation. Finally, the lumen in the realigned image was semi-manually segmented using SegEM.

\section{Results}

\subsection{Phantom}

\subsubsection{Micro CT} \label{offset_error}
\begin{figure}
	\centering
		\includegraphics[trim={0 0 0 0},clip,width=1\textwidth]{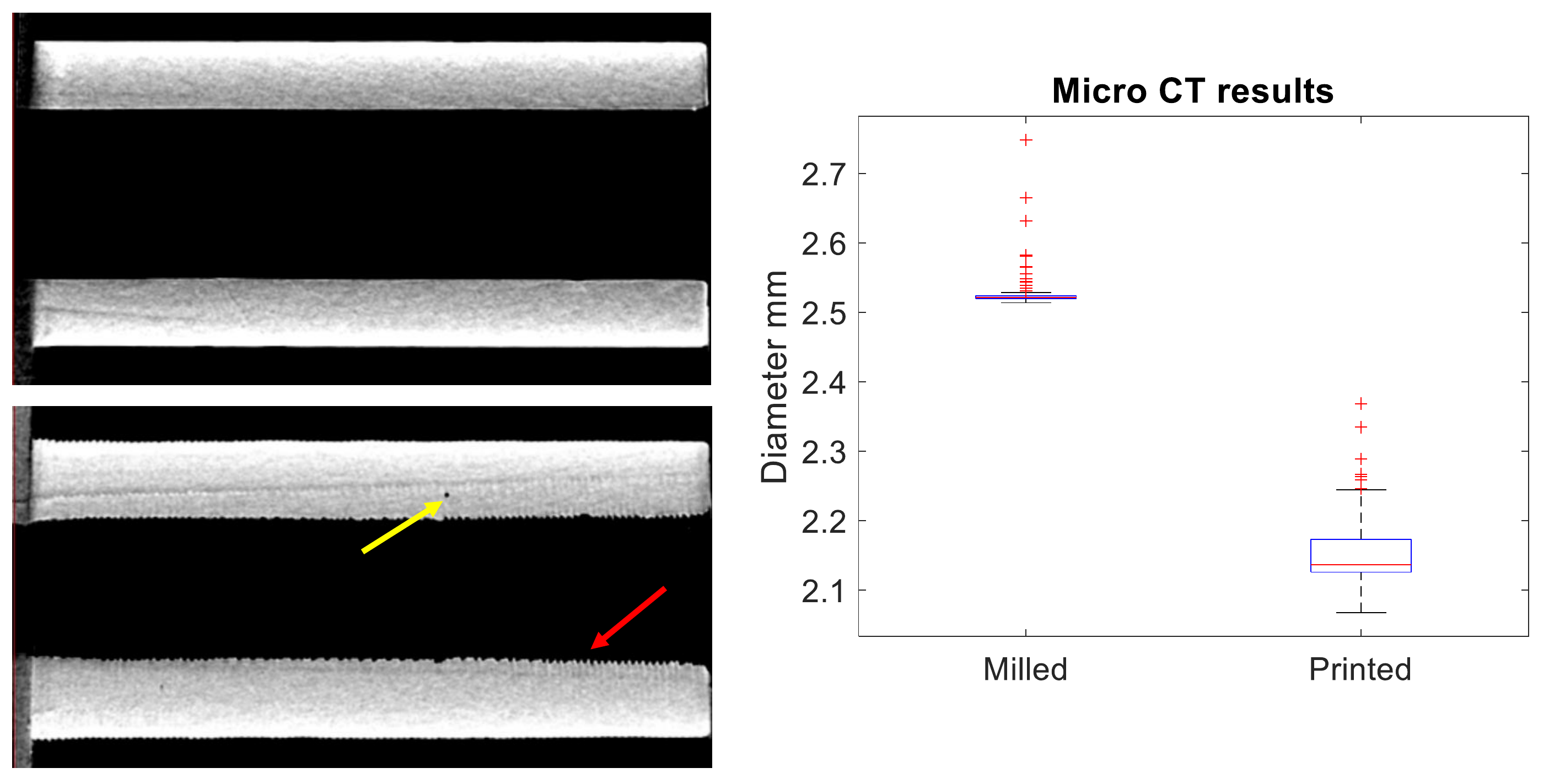}
	\caption{TOP LEFT: Slice of the 2.5mm diameter lumen made from a lathe. BOTTOM LEFT: Slice of the 2.5mm diameter lumen made using the 3D printer. The yellow arrow shows a hole within the resin. The red arrow shows the jagged surface. RIGHT: A box plot comparison of diameters between the milled and 3D printed lumens.}
	\label{Mirco_CT}
\end{figure}

The micro CT scanned images of the milled and 3D printed lumens are displayed in Figure \ref{Mirco_CT}. The 3D printed lumen contained abnormalities in the structure. Errors include holes and jagged surface indicated by the yellow and red arrows respectively.

Diameter measurements were taken 449 times in each slice from the tip to the base of the lumen. Figure \ref{Mirco_CT}, shows a systematic underestimation of the 3D printed lumen compared to the milled counterpart. The outliers were located near the ends of the lumen. The outliers were removed by only considering two-thirds of the lumen starting at the midpoint and expanding in both directions. We define mean difference between the inner lumen of the milled and 3D printed diameters as the offset error, and was calculated as 0.38mm.


\subsubsection{Clinical CT}
We present the diameter measurements from each of the 3D printed structures. However, we first corrected the offset error found in Section \ref{offset_error}. To this end, we added 0.38mm to every diameter measurement from the clinical CT phantom.

\begin{figure}
	\centering
		\includegraphics[trim={0 0 0 0},clip,width=1\textwidth]{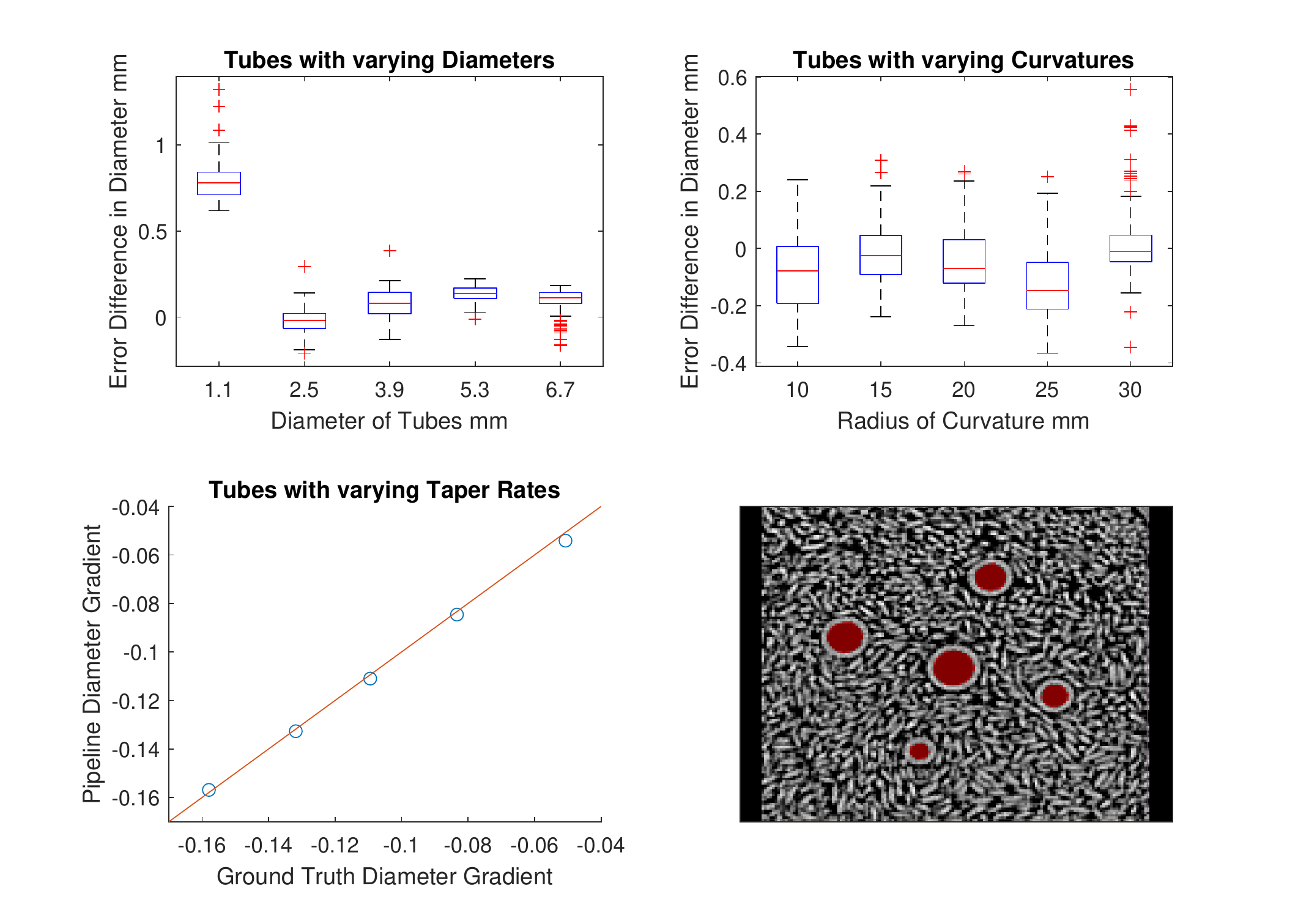}
	\caption{TOP LEFT: The diameter error difference when measuring the five tubes of varying diameters. TOP RIGHT: The diameter error difference when measuring the five tubes of varying curvatures. All tubes have a diameter of 2.5mm. BOTTOM LEFT: Relationship between the computed diameter gradient and its corresponding ground truth on the tapering lumen with the identity line. BOTTOM RIGHT: A slice of the tapering lumen, with the segmentation indicated in red.}
	\label{Finalised_graph_on_phantom}
\end{figure}

For the differing diameter lumen, Figure \ref{Finalised_graph_on_phantom} shows the accuracy was at sub voxel level for diameters above and equal to 2.5mm. In the 1.1mm lumen, there is a significant overestimation by approximately 0.8mm. In terms of precision, the error range for each lumen was within 0.3mm and therefore at sub voxel level. When the measured diameters were plotted against the ground truth without calibration, there is an offset of -0.44mm in the line of best fit.

For the differing curvature lumen, Figure \ref{Finalised_graph_on_phantom} shows the pipeline is accurate to sub voxel level. The mean error difference was within 0.2mm and the interquartile range was within 0.2mm for all curvatures. Thus the pipeline is precise to the sub voxel level and the precision is independent of the orientation of the lumen.

For the differing taper lumen, we computed the diameter gradient from the image. The diameter measurements were plotted against the centreline length of the tube, starting from the base. Next, the diameter gradient was taken as the gradient of the linear regression of the plotted data. Figure \ref{Finalised_graph_on_phantom}, shows good correspondence between the measured and ground truth diameter gradient. The correlation coefficient was  $r > 0.99$.


\subsection{Clinical Images}

\subsubsection{Tapering measurement}
Figure \ref{Method_ind_airway}, demonstrates the good linear fit between the arc length and the log cross sectional area. The tapering measurement shows a statistical difference between radiological normal and diseased airways, described in Figure \ref{Finalised_clincal_on_image}. On a Wilcoxon Rank Sum Test between the populations, $p = 7.1 \times 10^{-7}$.

\subsubsection{Bifurcations}
\begin{figure}
	\centering
		\includegraphics[width=1\textwidth]{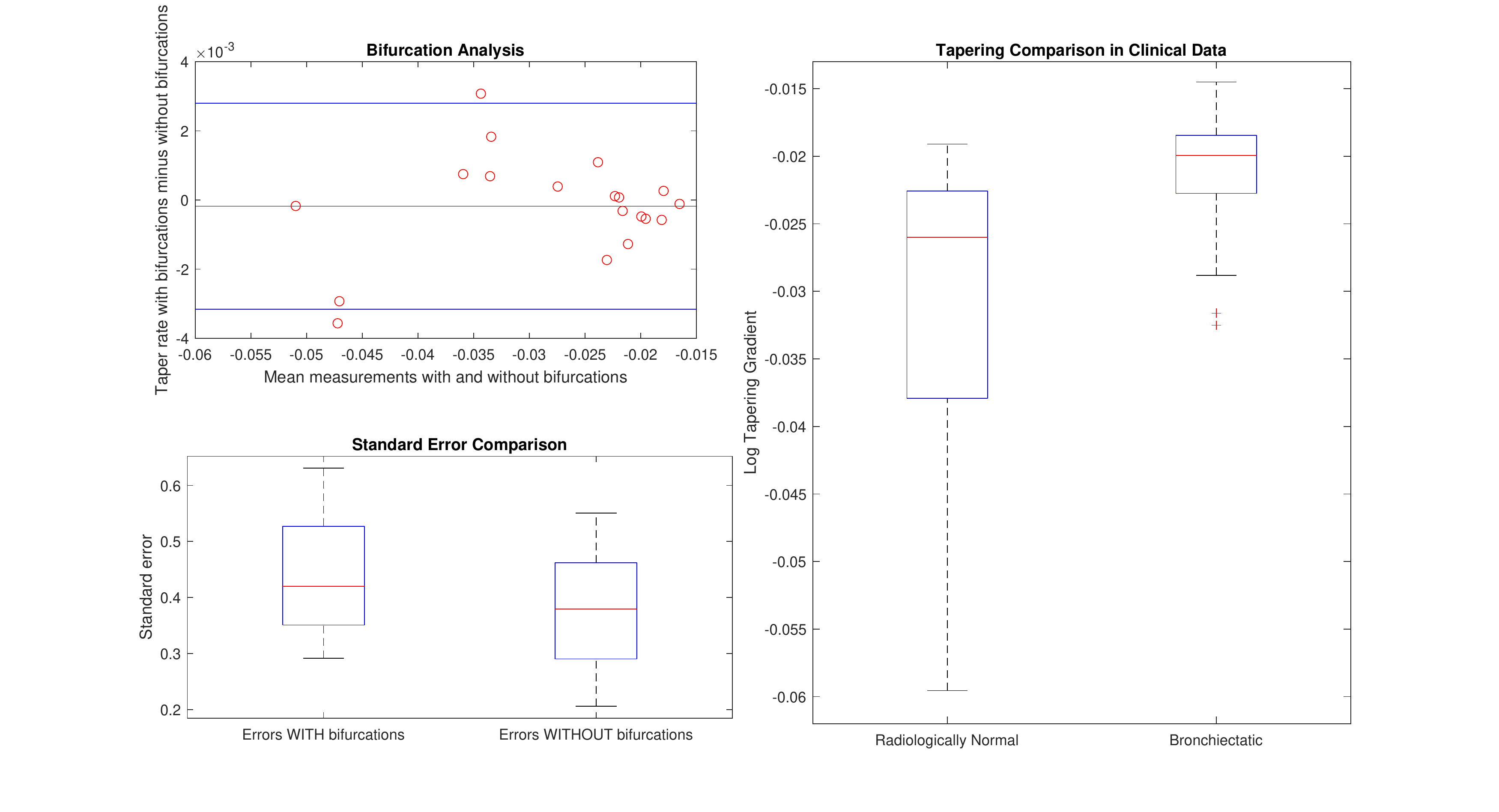}
	\caption{TOP LEFT: Bland-Altman graph\cite{Bland1986} showing the relationship of the taper rates $(n=19)$ with and without bifurcations. The black line indicates the mean $m$. The blue lines indicates $m \pm 1.96SD$, where $SD$ is the standard deviation. BOTTOM LEFT: Comparison of the standard error from linear regression between airways with and without bifurcations. On a Wilcoxon Rank Sum Test between the populations, $p = 3.4 \times 10^{-4}$. RIGHT: The collection of the log gradient for both the healthy $(n=35)$ and diseased $(n=39)$ airways. On a Wilcoxon Rank Sum Test between the populations, $p = 7.1 \times 10^{-7}$.}
	\label{Finalised_clincal_on_image}
\end{figure}

Figure \ref{Finalised_clincal_on_image}, shows a good correspondence in taper values with and without bifurcations, with correlation coefficient of $r=0.99$. We analyse the uncertainty of tapering measurement by computing the standard error of estimate\cite{Spiegal1998} on each linear regression between log cross sectional area and arc length. From Figure \ref{Finalised_clincal_on_image} there is a decrease in the error estimate when bifurcations were removed. A statistical difference was found between the error estimate with and without bifurcations. On a Wilcoxon Rank Sum Test between the populations, $p = 3.4 \times 10^{-4}$.


\section{Discussion and Conclusions}
The contribution of this paper was a pipeline to measure the airway taper rate. The tapering measurement is the gradient of the linear regression between the log cross-sectional area and arc length. This is equivalent to an exponential decay constant. Furthermore, using the taper measurement on clinical chest CT, we quantified different taper rates between bronchiectatic and radiologically normal airways. 

The paper presents a set of experiments to quantify the accuracy and precision of the pipeline. To this end, we developed a phantom with calibrated 3D printed structures. Using the phantom we showed the pipeline can measure diameters accurate to 0.3mm -- independent of orientation. As expected, the accuracy was reduced when the size of the lumen was near the voxel size of the image. From Figure \ref{Finalised_graph_on_phantom}, the pipeline can measure a range of linear changes in diameter along the centreline of the vessel. We demonstrated the taper measurement was robust against bifurcations. Figure \ref{Finalised_clincal_on_image}, showed the largest tapering difference caused by bifurcation was 0.0036 -- a small fraction in the range of possible taper values. 

In this paper, we showed that the tapering measurement gives promising clinical results. When the pipeline was applied to a group of bronchiectatic and healthy airways, the tapering rate of diseased airways was statistically different to controls. The diseased airways demonstrated a reduction in taper rate thereby indicating that our measurement can infer that parts of the airway are dilated.

The main disadvantage of the pipeline was the long processing time. Each airway requires a distal point, complete segmentation and thousands of points to be interpolated. The entire process can take hours to compute. The most exhaustive process was manually extending the airway segmentation to the most distal point. This was necessary as the airway segmentation was unable to automatically segment the required region. 

In conclusion, the proposed tapering measurement shows promising results towards distinguishing between bronchiectatic and healthy airways. In future work, we will further analyse the precision and stability of the pipeline; investigating the influence of noise and reconstruction algorithms. Finally, we will investigate the utility of our tapering measurements in terms of monitoring the changes of airways as the disease progress.


\section{Acknowledgements}
This work is supported by the EPSRC-funded UCL Centre for Doctoral Training in Medical Imaging (EP/ L016478/1) and the Department of Health’s NIHR-funded Biomedical Research Centre at University College London Hospitals.

KQ would like to thank Sara Reis and  Bjoern Eiben for their helpful discussions for writing this paper. In addition, KQ would like to thank Felix J S Bragman and Tom Doel for helpful discussions in implementing the pipeline.

\bibliography{report} 
\bibliographystyle{plain} 

\end{document}